# Sparse Data Driven Mesh Deformation

LIN GAO, Institute of Computing Technology, Chinese Academy of Sciences
YU-KUN LAI, School of Computer Science & Informatics, Cardiff University
JIE YANG, Institute of Computing Technology, Chinese Academy of Sciences
LING-XIAO ZHANG, Institute of Computing Technology, Chinese Academy of Sciences
LEIF KOBBELT, RWTH Aachen University
SHIHONG XIA, Institute of Computing Technology, Chinese Academy of Sciences

Example-based mesh deformation methods are powerful tools for realistic shape editing. However, existing techniques typically combine all the example deformation modes, which can lead to overfitting, i.e. using an overly complicated model to explain the user-specified deformation. This leads to implausible or unstable deformation results, including unexpected global changes outside the region of interest. To address this fundamental limitation, we propose a sparse blending method that automatically selects a smaller number of deformation modes to compactly describe the desired deformation. This along with a suitably chosen deformation basis including spatially localized deformation modes leads to significant advantages, including more meaningful, reliable, and efficient deformations because fewer and localized deformation modes are applied. To cope with large rotations, we develop a simple but effective representation based on polar decomposition of deformation gradients, which resolves the ambiguity of large global rotations using an as-consistent-as-possible global optimization. This simple representation has a closed form solution for derivatives, making it efficient for sparse localized representation and thus ensuring interactive performance. Experimental results show that our method outperforms state-of-the-art data-driven mesh deformation methods, for both quality of results and efficiency.

## 1 INTRODUCTION

Mesh deformation is a fundamental technique for geometric modeling. It has wide applicability ranging from shape design to computer animation. For pose changes of articulated objects, e.g. human bodies, deformation can be modeled using skeletons, although extra effort is needed to construct skeletons, and it is not suitable for deformation of general shapes. For improved flexibility, cage-based deformation resorts to cages that enclose the mesh as proxies (e.g. [Ju et al. 2008]). However, effort is also needed to build cages, and it requires experience to manipulate cages to obtain desired deformation.

Surface based methods allow general surface deformation to be obtained with an intuitive user interface. Typically, the user can specify a few handles on the given mesh, and by moving the handles to new locations, the mesh is deformed accordingly. Geometry based methods produce deformed surfaces following user constraints by optimizing some geometry related energies (e.g. [Levi and Gotsman 2015; Sorkine and Alexa 2007]). However, real-world deformable objects have complex internal structures, material properties and behavior which cannot be captured by geometry alone. As a result, such methods either require a large number of constraints or do not produce desired deformation for complex scenarios.

The idea of data-driven shape deformation is to provide explicit examples of how the input shape should look like under some example deformations (example poses) and then to interpolate between these poses in order to obtain a specific shape/pose instead of using synthetic basis functions or variational principles to drive the deformation. From the data interpolation perspective, we can consider each example deformation as a sample in a very high dimensional space (e.g. three times the number of vertices dimensional). Obviously, not all coordinates in this high dimensional space represent meaningful shapes. In fact, the effective shape space, i.e. the set of all *meaningful* deformations forms a relatively low dimensional sub-manifold in the high dimensional space of all *possible* deformations.

While in some papers [Heeren et al. 2014] the shape space is modeled mathematically, the data-driven approach reduces to a sophisticated weighted blending of the input poses. The existing methods in this area differ in how they represent deformations, i.e., how they encode a deformation by some high dimensional feature vector. The implicit underlying assumption for the blending operation is that any shape in the convex hull of the example deformations is meaningful. This, however, is not true in many cases. The most intuitive morphing path from one shape to another is not straight but follows a geodesic path on the shape space manifold as explored in [Heeren et al. 2014; Kilian et al. 2007].

This work was supported by the National Natural Science Foundation of China (No. 61502453 and No. 61611130215), Royal Society-Newton Mobility Grant (No. IE150731), CCF-Tencent Open Research Fund (No. AGR20160118), Knowledge Innovation Program of the Institute of Computing Technology of the Chinese Academy of Sciences (ICT20166040).

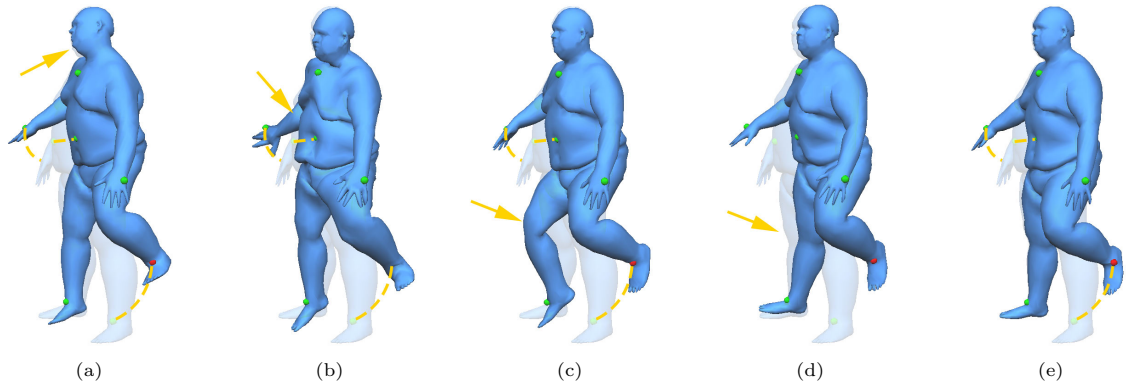

(a)　　　(b)　　　(c)　　　(d)　　　(e)

Fig. 1. Our interactive modeling technique deforms meshes by blending deformation modes derived from a number of input meshes. This data-driven approach avoids complex mathematical deformation models/energies and ensures plausibility of the deformation. While existing methods like (a) [Sumner et al. 2005], (b) [Fröhlich and Botsch 2011], and (c) [Gao et al. 2016] use globally supported deformation modes and always compute a blend of all available modes, we use deformation modes with local support (d) and only blend the most relevant ones (e) for a user-specified shape deformation. The sparsity of our method effectively avoids artifacts and side-effects that other methods produce, like unintended changes in regions that should not be affected by the edit (see yellow arrows).

In this paper, given a set of basis deformation modes, we propose *sparsity* as an effective mechanism to select a compact subset of suitable deformation modes, which when blended effectively satisfy the boundary conditions (i.e. handle movement). By using fewer basis modes, the deformed models are more likely to stay on the manifold. Moreover, it avoids overfitting and produces more robust deformation results. The basis deformation modes can have a variety of sources: example shapes themselves, as well as global and/or local deformation modes extracted from examples. Our method is able to choose suitable basis deformations, including suitable scales when multiscale deformation modes are provided, to produce meaningful results.

To represent large-scale deformations effectively, we further propose a simple and effective deformation representation which is able to cope with large-scale rotations. The ability to represent large-scale deformations is essential for data-driven deformation, since to fully exploit latent knowledge of example shapes, it is often needed to blend shapes with relatively large rotations, such as movement of the tail of a cat, and movement of hands (where deformation is driven by shoulder and elbow joints). The representation is well defined using an as-consistent-as-possible global optimization and has a closed form solution for derivatives, allowing efficient optimization for deformation. Our method has significant advantages, including realistic and controllable deformation avoiding side effects because suitable basis deformations are chosen automatically, and being much more efficient due to our deformation representation and precomputation, making it possible to exploit substantially larger basis than state-of-the-art methods while still performing at interactive rates. Figure 1 shows an example of data-driven deformation using the MPI DYNA dataset [Pons-Moll et al. 2015] with shapes of one subject as examples. Note that our algorithm is a general example-based deformation method. In addition to articulated deformation, the human example in Fig. 1 also involves non-articulated deformation (e.g. the tummy); more non-rigid deformations of general shapes will be shown later in the paper. Compared with state-of-the-art data-driven deformation methods [Fröhlich and Botsch 2011; Gao et al. 2016; Sumner et al. 2005], our method avoids unexpected deformation (e.g. the movement of head in (a) and the movement of right foot in (a-c)). Our sparsity regularization term ensures the unedited regions remain unchanged (e), which is not achieved without this regularization (d).

Our contributions are threefold:

- We propose a novel embedding (encoding) of deformations of triangle meshes that can handle arbitrarily large rotations in a stable way.
- We introduce sparsity in the weighted shape blending operator which leads to more plausible deformations.
- We present a highly efficient scheme to compute the involved operations in realtime. This is achieved by pre-computing all terms that do not depend on the position of the handles that the user moves to control the shape.

In Sec. 2, we review the most related work. We then give the detailed description of our novel mesh deformation representation in Sec. 3, followed by sparse data-driven deformation in Sec. 4. We present experimental results, including extensive comparisons with state-of-the-art methods in Sec. 5. Finally we draw conclusions in Sec. 6.

## 2 RELATED WORK

Mesh deformation has received significant attention. A comprehensive survey is beyond the scope of this paper. For interested readers, please refer to [Botsch and Sorkine 2008;

Gain and Bechmann 2008]. We review the work most related to ours.

**Geometry-based mesh deformation.** Geometry-based methods formulate mesh deformation as a constrained optimization problem with user specified handles and their locations as constraints. Terzopoulos et al. [1987] optimize a shell energy to obtain deformed meshes. Kobbelt et al. [1998] firstly propose a multi-resolution Laplacian-based deformation method. Sorkine et al. [2004] deform surfaces by minimizing differences of Laplacian coordinates before and after deformation. Since the representation is not rotation invariant, heuristics are needed to estimate the deformed Laplacian coordinates, which can be inaccurate. Yu et al. [2004] obtain deformed meshes by interpolating gradient fields derived from user constraints, fused using Poisson reconstruction. Iterative dual Laplacian [Kin-Chung Au et al. 2006] and volumetric graph Laplacian [Zhou et al. 2005] are also proposed to improve rotation and volume preservation. Sorkine and Alexa [2007] develop a mesh deformation method based on minimizing an as-rigid-as-possible (ARAP) energy, which measures non-rigid distortions in local 1-ring neighborhoods of each vertex. This is further improved by Levi and Gotsman [2015] with enhancement of smooth rotations (SR-ARAP). Such ARAP based methods can cope with large rotations well. However, all the geometry based methods do not capture the deformation behavior of the objects, so can produce undesirable results, especially for complicated objects and large-scale deformations. Physical principles such as modal analysis are also employed for interactive shape editing [Hildebrandt et al. 2011].

**Data-driven mesh deformation.** To address the limitations of geometry-based methods, data-driven methods learn typical deformations from examples and thus can produce more realistic deformation results. Sumner et al. [2005] propose a method for mesh deformation by blending deformation gradients of example shapes. The method is able to handle rotations, but fails to produce correct results for large rotations (of more than 180°). Fröhlich and Botsch [2011] instead use rotation invariant quantities, namely edge lengths, dihedral angles and volumes, to represent deformed meshes. The method is effective in interpolating (blending) example shapes, but does not handle extrapolation well, as this may lead to negative edge lengths. Gao et al. [2016] develop a method based on blending rotation differences between adjacent vertices and scaling/shear at each vertex. As rotation differences cancel out global rotations, the representation is rotation invariant. The method is able to handle both interpolation and extrapolation, thus produces improved deformation results. However, all of these methods are based on *global* blending of *all* the basis modes extracted from examples. Therefore, they tend to overfit and introduce unintended deformations, e.g. in areas far from user constraints during local fine-tuning. They also tend to use a large number of basis modes to represent even relatively local editing, and can produce deformation sequences which are not smooth with sudden changes when different sets of basis modes are chosen. Such methods [Gao et al. 2016; Sumner et al. 2005] use Principal Component Analysis (PCA) to analyze example datasets which can help reduce basis modes, but they still have similar drawbacks since all the modes are still used.

For articulated mesh models, Lewis et al. [2000] augment the Skeletal Subspace Deformation (SSD) method with displacements which are obtained by interpolating exemplar models created by artists. Weber et al. [2007] propose a skeleton based data driven deformation method for articulated mesh models, which works well with a small number of exemplar models. Sloan et al. [2001] provide a shape modeling system by interpolation of exemplar models using a combination of linear and radial basis functions. Unlike such methods, our proposed representation handles general deformations beyond articulation.

**Sparsity based geometric modeling.** The pioneer work by Neumann et al. [2013] proposes to use spatially localized basis to represent deformations. However, they represent shapes in the Euclidean coordinates, which are translation and rotation sensitive. Hence the method cannot handle large-scale deformations well. Huang et al. [2014] extend [Neumann et al. 2013] with a deformation gradient representation to handle larger rotations. However, this work cannot cope with rotations larger than 180° due to the limitation of deformation gradients. Wang et al. [2016a] extend [Neumann et al. 2013] using rotation invariant features based on [Fröhlich and Botsch 2011], and hence has a similar limitation when extrapolation of examples is required. Different from these works [Huang et al. 2014; Neumann et al. 2013; Wang et al. 2016a] which explore sparsity to localize deformation components, our method instead introduces sparse promoting regularization which prefers fewer basis modes for representing deformed shapes. This ensures suitable compact basis modes are chosen, which helps produce more meaningful deformation. When combined with sparse localized basis, our method avoids unwanted global deformations. Our method is also substantially faster than existing methods, thanks to our deformation representation and precomputation.

Sparsity has shown to be an effective tool for a variety of geometric modeling and processing tasks, including smooth skinning decomposition for skeleton-based deformation [Le and Deng 2012, 2013], non data driven shape deformation and editing [Deng et al. 2013; Gao et al. 2012], feature-preserving surface denoising [Wang et al. 2014], surface reconstruction using sparse dictionary learning [Xiong et al. 2014], non-rigid registration [Yang et al. 2015], surface approximation via redundant basis functions and parameterization optimization [Xu et al. 2016], and manifold approximation with redundant atom functions [Wang et al. 2016b]. Please refer to [Xu et al. 2015] for a recent survey.

Recent work [Wampler 2016] generalizes as-rigid-as-possible deformation [Sorkine and Alexa 2007] to take multiple examples and allow contributions of individual examples to vary spatially over the deformed surface. This helps make the method more descriptive. However, fundamental limitations such as potential non-local deformation effect and high complexity with large number of examples remain. The method uses convex weights summed to one, which has side effect of sparsity. However, sparsity is not directly optimized, nor the effect studied in their paper. In comparison, we introduce the sparsity regularization *explicitly* and on purpose, and our method is also able to handle shapes with large rotations especially when extrapolation is involved which are not supported by [Wampler 2016].

**Deformation representation.** Deformation representation is important for effective data-driven deformation. Euclidean coordinates are the most straightforward way [Bernard et al. 2016; Neumann et al. 2013], although with obvious limitations for rotations. More effective deformation gradients are used to represent shape deformations [Huang et al. 2014; Sumner et al. 2005], which still cannot handle large rotations. Another stream of research considers local frame based rotation-invariant representations [Baran et al. 2009; Kircher and Garland 2008; Lipman et al. 2005] which are capable of representing shapes with large rotations. However, such methods need to specify local frames compatibly with handle positions as constraints. They are suitable for mesh interpolation, but for mesh deformation, they require the user to specify not only positional constraints but also orientations of local frames compatibly, which can be a difficult task, making them unsuitable for mesh deformation.

As an improvement, Gao et al. [2016] use rotation differences to represent deforming shapes which can deform shapes with only positional constraints. It has a main drawback that the energy function to reconstruct surfaces from the representation is complicated, with no closed form derivative formulation. As a result, they resort to numerical derivatives, which makes it expensive to optimize the deformation energy. To address such limitations, we propose a new deformation representation, which is able to cope with large rotations. It also has significant advantages: it is efficient to compute and its derivatives have closed form solutions.

## 3 DEFORMATION REPRESENTATION

We first introduce the formulation of our deformation representation, and describe a simple algorithm to calculate the representation given a deformed mesh. Finally, we convert the deformation representation into the form of a feature vector.

Our method starts from deformation gradients, which are widely used in geometric modeling. Given a set of singly connected deforming shapes $S_k$, $k \in [1, \dots, n]$, where $n$ is the number of shapes. We assume that they have the same connectivity, which is often the case for shape datasets and can be obtained by consistent remeshing. Without loss of generality, we select the first mesh model as the reference model. Denote by $\mathbf{p}_{k,i} \in \mathbb{R}^3$ the $i^{\text{th}}$ vertex on the $k^{\text{th}}$ mesh. Then deformation gradient $\mathbf{T}_{k,i} \in \mathbb{R}^{3 \times 3}$ defined on the 1-ring neighborhood of $\mathbf{p}_{k,i}$ can be calculated by the following least squares formulation:

$$\arg \min_{\mathbf{T}_{k,i}} \sum_{j \in N(i)} c_{ij} \|(\mathbf{p}_{k,i} - \mathbf{p}_{k,j}) - \mathbf{T}_{k,i}(\mathbf{p}_{1,i} - \mathbf{p}_{1,j})\|_2^2. \quad (1)$$

$N(i)$ is the neighborhood vertex index set of the $i^{\text{th}}$ vertex. $c_{ij} = \cot \alpha_{ij} + \cot \beta_{ij}$ is the cotangent weight, where $\alpha_{ij}$ and $\beta_{ij}$ are angles in the adjacent triangles opposite to the edge $(i, j)$. This is used to avoid discretization bias in deformation [Gao et al. 2016; Levi and Gotsman 2015]. For the rank-deficient case, we add the normal of the plane to the 1-ring edges for computing the deformation gradient to ensure a unique solution.

The main drawback of the deformation gradient representation is that it cannot handle large-scale rotations. By polar decomposition, the deformation gradient $\mathbf{T}_{k,i}$ can be decomposed into a rigid rotation matrix $\mathbf{R}_{k,i}$ and a scale/shear matrix $\mathbf{S}_{k,i}$: $\mathbf{T}_{k,i} = \mathbf{R}_{k,i}\mathbf{S}_{k,i}$. The scale/shear matrix $\mathbf{S}_{k,i}$ is uniquely defined. However, given the rigid rotation $\mathbf{R}$, there are infinite corresponding rotation angles. To ensure uniqueness, typical formulations constrain the rotation angles to be within $[0, \pi]$ which are unsuitable for smooth large-scale deformations.

In order to handle large-scale rotations, we take the axis-angle representation to represent the rotation matrix $\mathbf{R}_{k,i}$. The rotation matrix $\mathbf{R}_{k,i}$ can be represented using a rotation axis $\boldsymbol{\omega}_{k,i}$ and rotation angle $\theta_{k,i}$ pair with the mapping $\phi$:

$$\phi(\boldsymbol{\omega}_{k,i}, \theta_{k,i}) = \mathbf{R}_{k,i}, \quad (2)$$

where $\boldsymbol{\omega}_{k,i} \in \mathbb{R}^3$ and $\|\boldsymbol{\omega}_{k,i}\| = 1$. Given $\mathbf{R}_{k,i}$, assuming $\phi(\boldsymbol{\omega}_{k,i}, \theta_{k,i}) = \mathbf{R}_{k,i}$ is an equivalent representation, then any representation in the set

$$\Omega_{k,i} = \{(\boldsymbol{\omega}_{k,i}, \theta_{k,i} + t \cdot 2\pi), (-\boldsymbol{\omega}_{k,i}, -\theta_{k,i} + t \cdot 2\pi)\} \quad (3)$$

is also a possible value, where $t$ is an arbitrary integer.

*As-consistent-as-possible deformation representation.* In order to handle large-scale rotations, the orientations of rotation axes and rotation angles of adjacent vertices need to be as consistent as possible. For 2D deformation, some pioneer work [Baxter et al. 2008; Jaeil and Andrzej 2003] exploits a similar idea to consistently set rotation angles of vertices to deal with large-scale deformation. However, 3D deformation is much more challenging. Instead of using a greedy approach as these 2D methods, we model this problem using an as-consistent-as-possible principle and formulate the optimization problem as one that maximizes the overall consistency. More specifically, the consistency for deformation means that the difference of rotation angles and the angle between rotation axes should be as small as possible.

We first consider consistent orientation for axes. Assuming $\boldsymbol{\omega}_{k,i}$ is an arbitrarily oriented axis direction for the $i^{\text{th}}$ vertex on the $k^{\text{th}}$ mesh. Denote by $o_{k,i}$ a scalar indicating potential orientation reversal of the axis, where $o_{k,i} = 1$ (resp. $-1$) means the orientation of $\boldsymbol{\omega}_{k,i}$ is unchanged (resp. inverted). Then consistent orientation of axes is turned into a problem of finding a set of $o_{k,i}$ that maximizes the compatibility of axis orientations between adjacent vertices:

$$\arg\max_{o_{k,i}} \sum_{(i,j)\in\mathcal{E}} o_{k,i} o_{k,j} \cdot s(\boldsymbol{\omega}_{k,i} \cdot \boldsymbol{\omega}_{k,j}, \theta_{k,i}, \theta_{k,j}) \qquad (4)$$

$$\text{s.t.} \quad o_{k,1} = 1, o_{k,i} = \pm 1 (i \neq 1),$$

where $\mathcal{E}$ is the edge set, and $s(\cdot)$ is a function measuring orientation consistency, and is defined as follows:

$$s(\cdot) = \begin{cases} 0, & |\boldsymbol{\omega}_{k,i} \cdot \boldsymbol{\omega}_{k,j}| \leq \epsilon_1 \quad or \quad \theta_{k,i} < \epsilon_2 \quad or \quad \theta_{k,j} < \epsilon_2 \\ 1, & \text{Otherwise if } \boldsymbol{\omega}_{k,i} \cdot \boldsymbol{\omega}_{k,j} > \epsilon_1 \\ -1, & \text{Otherwise if } \boldsymbol{\omega}_{k,i} \cdot \boldsymbol{\omega}_{k,j} < -\epsilon_1 \end{cases}$$
$$(5)$$

The rationale of the definition above is based on the following cases: In general, when the angle between the rotation axes of adjacent vertices is less than $90°$ (resp. greater than $90°$), the function value is 1 (resp. $-1$), meaning such cases are preferred (resp. not preferred). However, there are two exceptions: If the axes are near-orthogonal ($\epsilon_1 = 10^{-6}$ in our experiments), the function value is set to 0, which improves the robustness to noise. Another situation is when one of the vertices has near zero rotation ($\epsilon_2 = 10^{-3}$ in our experiments), the axis for such vertices can be quite arbitrary, so we do not penalize inconsistent orientation in such cases. Note that $s(\cdot)$ can be precomputed, and only $o_{k,i}$ needs to be optimized.

After optimizing the orientation of rotation axes, the rotation angles of adjacent vertices also need to be optimized to keep consistency. Once the axis orientation is fixed, the rotation angle can differ by a multiple of $2\pi$. The optimization is defined as follows:

$$\arg\min_{t_{k,i}} \sum_{(i,j)\in\mathcal{E}} \|(t_{k,i} \cdot 2\pi + o_{k,i}\theta_{k,i}) - (t_{k,j} \cdot 2\pi + o_{k,j}\theta_{k,j})\|_2^2$$
$$(6)$$

$$\text{s.t.} \quad t_{k,i} \in \mathbb{Z}, \ t_{k,1} = 0,$$

where $t_{k,i}$ is the cycle number for the $i^{\text{th}}$ vertex of the $k^{\text{th}}$ model. Both integer optimization problems can be solved by the constrained integer solver (CoMISo) [Bommes et al. 2010] effectively. CoMISo optimizes constrained interger problems successively by solving relaxed problems in the real value domain. The rotation angle optimization in Eqn. 6 is in a positive definite quadratic form which is a convex optimization problem during each iteration. Therefore, this optimization is insensitive to the initial value. In contrast, the axis orientation optimization in Eqn. 4 is non-convex in each optimization iteration, so good initialization is useful to speed up the convergence of this optimization. We use a simple and effective greedy algorithm based on breadth-first search to generate the initial solution. The algorithm repeatedly accesses an unvisited vertex adjacent to a visited one and selects the orientation which (locally) minimizes Eqn. 4. This process repeats until all the vertices have their axes assigned.

The axis-angle representation is well defined, i.e. unique up to a global shift of multiples of $2\pi$ for $\theta$, and/or simultaneous negating of $\boldsymbol{\omega}$ and $\theta$ globally. Equivalently, the representation is unique once the axis-angle representation of one vertex is fixed. Without loss of generality, we choose the rotation angle of the first vertex in $[0, 2\pi)$ and orientation to be $+1$. In the result section, we will demonstrate that this method is effective to handle large-scale deformations and high-genus models, and is robust to noise.

*Feature Vector Representation.* Given the axis-angle representation, it is not suitable to blend directly, so we convert this axis $\boldsymbol{\omega}$ and angle $\theta$ to the matrix log representation:

$$\log \mathbf{R} = \theta \begin{pmatrix} 0 & -\omega_z & \omega_y \\ \omega_z & 0 & -\omega_x \\ -\omega_y & \omega_x & 0 \end{pmatrix} \qquad (7)$$

Due to the matrix symmetry, for the $k^{\text{th}}$ shape, we collect the upper triangular matrix of $\log \mathbf{R}_{k,i}$ excluding the diagonal elements as they are always zeros, and the upper triangular matrix of scaling/shear matrix $\mathbf{S}_{k,i}$ (including the diagonal elements) to form the deformation representation feature vector $\mathbf{f}_k$, where the transformation at each vertex is encoded using a 9-dimensional vector. Using logarithm of rotation matrices makes it possible to linearly combine the obtained feature vectors [Alexa 2002; Murray et al. 1994]. The rotation matrix $\mathbf{R}_{k,i}$ can be recovered by matrix exponential $\mathbf{R}_{k,i} = \exp(\log \mathbf{R}_{k,i})$, where $\log \mathbf{R}_{k,i}$ is part of the feature vector. The dimension of $\mathbf{f}$ is denoted as $m = 9|\mathcal{V}|$, where $|\mathcal{V}|$ is the number of vertices.

## 4 SPARSE DATA DRIVEN DEFORMATION

Given $n$ example shapes $S_k, k \in [1, \ldots, n]$, we can obtain $n$ feature vectors $\mathbf{F} = \{\mathbf{f}_k\}$ using the representation described in Sec. 3. We first extract a basis of deformation modes from the given examples. $\mathbf{C} = \{\mathbf{c}_k\}, \mathbf{c}_k \in \mathbb{R}^{9|\mathcal{V}|}, k \in [1, \ldots, K]$ in our representation space, where $K$ is the number of basis deformations, which can be specified by the user. To produce a deformed mesh, assuming $H$ is the handle set, for each $h \in H$, the user specifies its location to be $\mathbf{v}_h$. Our data-driven deformation aims to find a shape compactly represented as a linear combination of basis modes, while satisfying the given user specification as *hard constraints*.

### 4.1 Extraction of basis deformations

Our method can take basis deformations $\mathbf{C}$ from different sources. If global deformation is desired, we can take all the example shapes in our deformation representation, i.e. $\mathbf{F}$ as the basis, or when the number of examples is large, PCA-based dimensionality reduction of $\mathbf{F}$. When local editing is desired, we employ the method [Neumann et al. 2013] on our representation to obtain the basis deformations $\mathbf{C}$ by optimizing the following:

$$\min_{\mathbf{W},\mathbf{C}} \|\mathbf{F} - \mathbf{CW}\|_F^2 + \Omega(\mathbf{C}), \qquad \text{s.t.} \quad \max_{i\in[1,n]} (|\mathbf{W}_{k,i}|) = 1, \forall k,$$
$$(8)$$

where $\mathbf{F}_{m\times n}$ is the matrix with each column corresponding to the deformation feature representation of each example, $\mathbf{C}_{m\times K}$ is the basis deformation components to be extracted. $\mathbf{W}_{K\times n}$ is the weight matrix. The condition on $\mathbf{W}_{k,i}$ avoids trivial solutions with arbitrarily large $\mathbf{W}$ values and arbitrarily small $\mathbf{C}$ values. The locality term $\Omega(\mathbf{C})$ is defined similar to [Neumann et al. 2013]:

$$\Omega(\mathbf{C}) = \sum_{k=1}^{K} \sum_{i=1}^{m} \Lambda_{k,i} \|\mathbf{c}_k^{(i)}\|_2. \qquad (9)$$

As in [Neumann et al. 2013], $\Lambda_{k,i}$ is defined to linearly map the range of geodesic distances from the $k^{\text{th}}$ centroid sample vertex in a given range $[d_{min}, d_{max}]$ to $[0, 1]$ (with geodesic distance out of the range capped). $\mathbf{c}_k^{(i)}$ represents the local deformation for the $i^{\text{th}}$ vertex of the $k^{\text{th}}$ basis deformation. Please refer to [Neumann et al. 2013] for more details regarding parameter settings and implementation.

The major difference between our method and existing methods is to use the $\ell_{2,1}$ norm in our *deformation representation*, instead of the vertex displacement in the Euclidean coordinates [Neumann et al. 2013] or the deformation gradient [Huang et al. 2014]. With our representation, we can deal with datasets with large-scale deformations much more effectively. As demonstrated in Fig. 2, we apply [Huang et al. 2014; Neumann et al. 2013] to the SCAPE data set [Anguelov et al. 2005] with the first four components shown, and the limitation is clearly visible. Our method produces plausible localized basis deformations.

### 4.2 Sparse shape deformation formulation

To deform the deformed mesh, we formulate the deformation gradient of the deformed mesh as a linear combination of the basis deformations; similar linear blending operators have been employed in [Sumner et al. 2005; Weber et al. 2007]:

$$\mathbf{T}_i(\mathbf{w}) = \exp(\sum_{l=1}^{K} w_l \log \tilde{\mathbf{R}}_{l,i}) \sum_{l=1}^{K} w_l \tilde{\mathbf{S}}_{l,i}, \qquad (10)$$

where $\mathbf{w}$ is the combination weight vector, $\log \tilde{\mathbf{R}}_{l,i}$ and $\tilde{\mathbf{S}}_{l,i}$ are part of the $l^{\text{th}}$ basis $\mathbf{c}_l$.

The vertex positions $\mathbf{p}_i' \in \mathbb{R}^3$ of the deformed mesh can be calculated by minimizing the following energy:

$$E(\mathbf{p}_i', \mathbf{w}) = \sum_i \sum_{j \in N(i)} c_{ij} \|(\mathbf{p}_i' - \mathbf{p}_j') - \mathbf{T}_i(\mathbf{w})(\mathbf{p}_{1,i} - \mathbf{p}_{1,j})\|_2^2. \qquad (11)$$

For each vertex $h$ in the handle set $H$, its target vertex position $\mathbf{p}_h' = \mathbf{v}_h$ is specified by the user and does not change over the optimization. This formulation however only aims at choosing basis deformations that minimize non-rigid distortions, and more basis deformations than necessary may be chosen. To produce more natural and realistic deformation and avoid overfitting, we further introduce the sparse regularization term such that the solution will prefer to use fewer basis deformations if possible. This along with sparse localized basis means that local deformation tends to be represented using local basis only, thus avoiding the unexpected global effect with traditional methods. The resulting formula with

sparse regularization is given as follows:

$$\tilde{E}(\mathbf{p}_i', \mathbf{w}) = \sum_i \sum_{j \in N(i)} c_{ij} \|(\mathbf{p}_i' - \mathbf{p}_j') - \mathbf{T}_i(\mathbf{w})(\mathbf{p}_{1,i} - \mathbf{p}_{1,j})\|_2^2 + \lambda \|\mathbf{w}\|_1. \qquad (12)$$

$\lambda$ is a parameter to control the importance of the sparsity regularization. Except for the experiments for analyzing its effect, $\lambda = 0.5$ is used for all our experiments.

### 4.3 Algorithmic solution without sparsity regularization

To make it easier to follow, we first describe the algorithmic solution to the problem $E(\mathbf{p}_i', \mathbf{w})$ without the sparse regularization term $\|\mathbf{w}\|_1$. We use the Gauss-Newton method to solve the optimization. In each step, we will solve a least squares problem. We derive this procedure with the Taylor expansion:

$$\mathbf{T}_i(\mathbf{w}_t + \Delta \mathbf{w}_t) = \mathbf{T}_i(\mathbf{w}_t) + \sum_l \frac{\partial \mathbf{T}_i(\mathbf{w}_t)}{\partial w_{t,l}} \Delta w_{t,l}. \qquad (13)$$

$\mathbf{w}_t$ is the weights in the $t^{\text{th}}$ Gauss-Newton iteration. $\Delta \mathbf{w}_t = \mathbf{w}_{t+1} - \mathbf{w}_t$. The derivative $\frac{\partial \mathbf{T}_i(\mathbf{w}_t)}{\partial w_{t,l}}$ is defined as follows:

$$\frac{\partial \mathbf{T}_i(\mathbf{w}_t)}{\partial w_{t,l}} = \qquad (14)$$
$$\exp(\sum_l w_l \log \tilde{\mathbf{R}}_{l,i}) \log \tilde{\mathbf{R}}_{l,i} \sum_l w_{t,l} \tilde{\mathbf{S}}_{l,i} + \exp(\sum_l w_{t,l} \log \tilde{\mathbf{R}}_{l,i}) \tilde{\mathbf{S}}_{l,i}$$

For simplicity, we take the notation $\mathbf{e}_{ij} = \mathbf{p}_i - \mathbf{p}_j$. Eqn. (11) can be derived as:

$$E(\mathbf{p}_i', \mathbf{w}_{t+1}) \qquad (15)$$
$$= \sum_i \sum_{j \in N(i)} c_{ij} \|\mathbf{e}_{ij}' - (\mathbf{T}_i(\mathbf{w}_t) + \sum_l \frac{\partial \mathbf{T}_i(\mathbf{w})}{\partial w_{t,l}}(w_{t+1,l} - w_{t,l}))\mathbf{e}_{1,ij}\|^2$$

Eqn. (15) leads to a least squares problem in the following form:

$$\arg\min_{\mathbf{x}} \|\mathbf{A}\mathbf{x} - \mathbf{b}\|_2^2 \qquad (16)$$

which can be efficiently solved using a linear system. Here, $\mathbf{A} \in \mathbb{R}^{3|\mathcal{E}| \times (3|\mathcal{E}| + K)}$, $\mathbf{x} \in \mathbb{R}^{3|\mathcal{E}| + k}$, $\mathbf{b} \in \mathbf{R}^{3|\mathcal{E}|}$. The matrix $\mathbf{A}$ has the following form:

$$\begin{pmatrix} \mathbf{L} & & & -\mathbf{J}_1 \\ & \mathbf{L} & & -\mathbf{J}_2 \\ & & \mathbf{L} & -\mathbf{J}_3 \end{pmatrix} \qquad (17)$$

where $\mathbf{x} = [\mathbf{p}_x', \mathbf{p}_y', \mathbf{p}_z', \mathbf{w}_{t+1}]^T$. $\mathbf{L}$ is a matrix of size $|\mathcal{E}| \times |\mathcal{V}|$, where $|\mathcal{E}|$ and $|\mathcal{V}|$ are the number of half edges and vertices respectively. $\mathbf{L}$ is highly sparse, with only two non-zero elements in each row i.e. the cotangent weights $c_{ij}$ and $-c_{ij}$ for the row corresponding to edge $(i, j)$. Three copies of $L$ appear in the matrix (17), corresponding to $x$, $y$ and $z$ coordinates. $\mathbf{J}_i(i = 1, 2, 3) \in |\mathcal{E}| \times K$ is the product of the Jacobian matrix $\frac{\partial \mathbf{T}_i(\mathbf{w})}{\partial w_{t,l}}$ and $\mathbf{e}_{1,ij}$. Minimizing Eqn. (16) is equivalent to solving the following normal equation:

$$\mathbf{A}^T \mathbf{A} \mathbf{x} = \mathbf{A}^T \mathbf{b}. \tag{18}$$

The matrix $\mathbf{A}^T \mathbf{A}$ can be written as:

$$\begin{pmatrix} \mathbf{L}^T \mathbf{L} & & & -\mathbf{L}^T \mathbf{J}_1 \\ & \mathbf{L}^T \mathbf{L} & & -\mathbf{L}^T \mathbf{J}_2 \\ & & \mathbf{L}^T \mathbf{L} & -\mathbf{L}^T \mathbf{J}_3 \\ -\mathbf{J}_1^T \mathbf{L} & -\mathbf{J}_2^T \mathbf{L} & -\mathbf{J}_3^T \mathbf{L} & \sum_{i=1}^3 \mathbf{J}_i^T \mathbf{J}_i \end{pmatrix} \tag{19}$$

We compute Cholesky decomposition of $\mathbf{A}^T \mathbf{A}$: $\mathbf{V}^T \mathbf{V} = \mathbf{A}^T \mathbf{A}$, where the upper triangular matrix $\mathbf{V} \in \mathbb{R}^{(3|\mathcal{E}|+K) \times (3|\mathcal{E}|+K)}$ has the following structure:

$$\begin{pmatrix} \mathbf{U} & & & \mathbf{U}_1 \\ & \mathbf{U} & & \mathbf{U}_2 \\ & & \mathbf{U} & \mathbf{U}_3 \\ & & & \mathbf{U}_4 \end{pmatrix} \tag{20}$$

and the following equations hold:

$$\mathbf{U}^T \mathbf{U} = \mathbf{L}^T \mathbf{L} \tag{21}$$

$$\mathbf{U}^T \mathbf{U}_i = -\mathbf{L}^T \mathbf{J}_i \qquad i \in \{1, 2, 3\} \tag{22}$$

$$\mathbf{U}_4^T \mathbf{U}_4 = \sum_{i=1}^3 \mathbf{J}_i^T \mathbf{J}_i \tag{23}$$

The matrix $\mathbf{L}$ is constant during deformation once the handle set $H$ is fixed, so we precompute Cholesky decomposition $\mathbf{U}^T \mathbf{U} = \mathbf{L}^T \mathbf{L}$ before real-time deformation, $\mathbf{U} \in \mathbb{R}^{|\mathcal{V}| \times |\mathcal{V}|}$. During deformation, $\mathbf{U}_i \in \mathbb{R}^{|\mathcal{V}| \times K}$ can be efficiently calculated by back substitution. The most time-consuming operation is $\sum_{i=1}^3 \mathbf{J}_i^T \mathbf{J}_i$, because two dense matrices are multiplied. However, we develop an efficient technique to solve this problem with precomputation, because $\mathbf{J}$ only changes when $\mathbf{w}$ is changed. For a typical scenario, this reduces the deformation time from 190ms to 5ms (see the Appendix for details). After the above computation, we get the upper triangular matrix (20), which can be used to obtain the positions $\mathbf{p}'$ and the weight $\mathbf{w}$ efficiently, using back substitution.

### 4.4 Optimization of the sparse deformation formulation

We now consider the formula $\tilde{E}(\mathbf{p}'_i, \mathbf{w})$ with sparse regularization (Eqn. (12)). We similarly use the Gauss-Newton method as described in Sec. 4.3. To cope with the sparse term on the weights, we incorporate the Alternating Direction Method of Multipliers (ADMM) optimization [Boyd et al. 2011] into the iteration of Gauss-Newton optimization. Similar to Eqn. (16), $\tilde{E}(\mathbf{p}'_i, \mathbf{w})$ can be rewritten in the form of

$$\|\mathbf{A}\mathbf{x} - \mathbf{b}\|_2^2 + \lambda \|\mathbf{x}\|_1. \tag{24}$$

To solve this Lasso problem [Boyd et al. 2011], we employ ADMM as follows. $\mathbf{x}_0$ is the initial value for $\mathbf{x}$, $\mathbf{z}$ and $\mathbf{u}$ are two auxiliary vectors initialized as $\mathbf{0}$. $\mathbf{x}, \mathbf{z}, \mathbf{u} \in \mathbb{R}^{3|\mathcal{E}|+K}$. ADMM method is used to work out $\mathbf{x}$ that optimizes Eqn. (24) by optimizing the following subproblems alternately:

$$\mathbf{x}_{k+1} = (\mathbf{A}^T \mathbf{A} + \rho \mathbf{I})^{-1} \cdot (\mathbf{A}^T \mathbf{b} + \rho(\mathbf{z}_k - \mathbf{u}_k)) \tag{25}$$

$$\mathbf{z}_{k+1} = \mathrm{shrink}(\mathbf{x}_{k+1} + \mathbf{u}_{k+1}, \lambda/\rho) \tag{26}$$

$$\mathbf{u}_{k+1} = \mathbf{u}_k + \mathbf{x}_{k+1} - \mathbf{z}_{k+1}. \tag{27}$$

In the above formulas, $shrink(x, a)$ is a function defined as $shrink(x, a) = (a - x)_+ - (-a - x)_+$, where the function $(x)_+ = \max(0, x)$. This ADMM algorithm is solved very efficiently in each Gauss-Newton iteration, because the matrix $(\mathbf{A}^T \mathbf{A} + \rho \mathbf{I})$ is unchanged during ADMM optimization iterations. We use an approach similar to Sec. 4.3 with precomputation described in the Appendix, by replacing $\mathbf{A}^T \mathbf{A}$ with $(\mathbf{A}^T \mathbf{A} + \rho \mathbf{I})$. $\mathbf{x}_{k+1}$ is calculated very efficiently using back substitution. The pseudocode for sparse data driven deformation is given in Algorithm 1. The parameter $\rho = 0.5$ is used in our experiments. In practice, we perform the outer Gauss-Newton iterations for 4 times and the inner ADMM iterations for 4 times, which is sufficient to produce good deformation results.

---

**ALGORITHM 1:** Sparse Data Driven Deformation

---

**Input:** $K$ deformation modes analyzed e.g. using Eqn. (8)
**Input:** Deformation handle $H$.
**Output:** Deformed mesh model.
**Preprocessing:**
Precompute the upper triangular matrix $\mathbf{U}$ by Cholesky decomposition with Eqn. (21)
Precompute the unchanged terms in Eqns. (28) and (29)
**Real-Time Deformation:**
**for** each Gaussian-Newton iteration **do**
    Compute $\mathbf{U}_i$ $i \in \{1, 2, 3\}$ by back substitution with Eqn. (22)
    Compute $\sum_{i=1}^3 \mathbf{J}_i^T \mathbf{J}_i$ with Eqn. (31)
    Compute $\mathbf{U}_4$ by Cholesky decomposition with Eqn. (23)
    Initialize $\mathbf{x}_0$, $\mathbf{z}_0$ and $\mathbf{u}_0$, set $k = 0$
    **for** each ADMM optimization iteration **do**
        Set $k = k + 1$
        Calculate $\mathbf{x}_{k+1}$ by back substitution with Eqn. (25)
        Calculate auxiliary variables $\mathbf{z}_{k+1}$ and $\mathbf{u}_{k+1}$ with Eqns. (26) and (27)
    **end for**
**end for**

---

## 5 EXPERIMENTAL RESULTS

Our experiments were carried out on a computer with an Intel-i7 6850K and 16GB RAM. Our code is CPU based, carefully optimized with multi-threading, which will be made available to the research community. We use both synthetic shapes and shapes from existing datasets, including the SCAPE dataset [Anguelov et al. 2005], and the MPI Dyna dataset [Pons-Moll et al. 2015]. We simplify the SCAPE dataset to $4K$ triangles and Dyna dataset to $6K$ triangles using [Garland and Heckbert 1997], as this provides faster response while ensuring deformation quality.

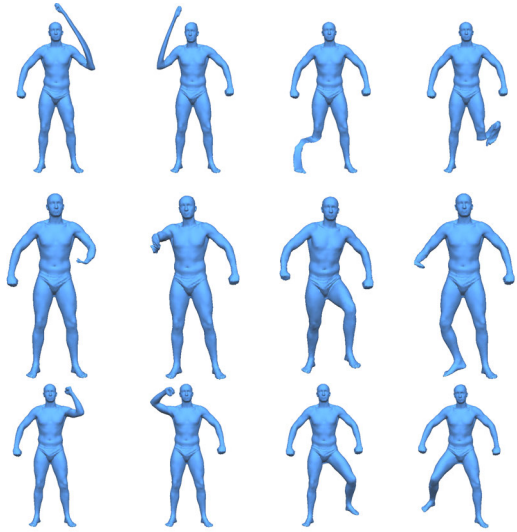

Fig. 2. Comparison of different localized principal component analysis methods, with the first four principal components of each method shown. First row: results of [Neumann et al. 2013], second row: results of [Huang et al. 2014], third row: results of our method. The methods [Huang et al. 2014; Neumann et al. 2013] cannot cope with large rotations, resulting in distorted basis deformations. In contrast, our method capable of handling large rotations produces reasonable basis deformations.

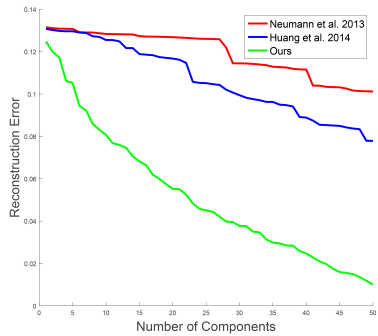

Fig. 3. Generalization of basis to new shapes. We plot the reconstruction error ($y$-axis) with respect to the number of components used ($x$-axis) on a set of human shapes from [Anguelov et al. 2005] that are not part of the training data. Our method outperforms alternative methods with significantly lower reconstruction errors.

To produce deformation results with more details, we also implemented multiresolution optimization similar to [Gao et al. 2016]. As shown in Fig. 27, the original model is with 13700 triangles, and the simplified coarse mesh contains 6528 triangles. With 30 deformation modes, our data-driven deformation method takes 117ms on the coarse mesh, and takes another 12ms to drive the original dense mesh, so in total it takes 129ms to achieve rich deformation with fine details. For other cases, meshes with 4K-6K triangles produce visually similar results as with high resolution meshes.

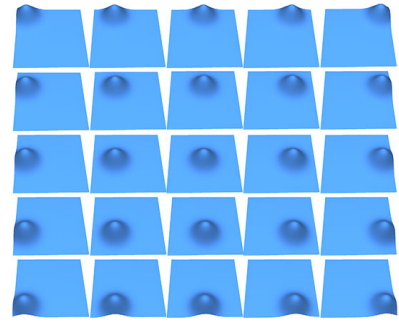

Fig. 4. A synthetic basis containing 25 deformation modes, each is a deformed square with a 2D Gaussian distribution offset, centered in a regular grid.

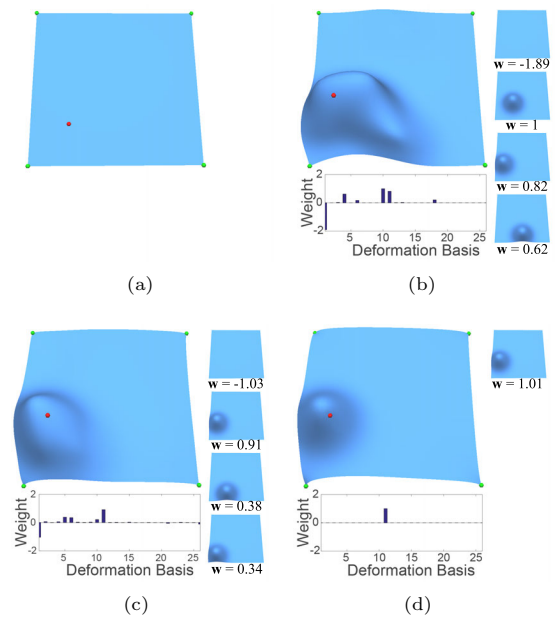

Fig. 5. Deformation results with the basis from Fig. 4 using different methods with the same control point trajectory. The green handles are fixed while the red handle is being moved. (a) input shape, (b) result of [Gao et al. 2016], (c) result of our optimization without the sparsity term, (d) our deformation result with the sparsity term. We also visualize the dominant basis modes and their contribution weights. With sparsity regularization, our method produces plausible deformation with smaller number of basis modes involved, which ensures the resulting deformation is closer to the examples.

**Evaluation of localized basis extraction.** We analyze the localized basis extracted using our rotation representation. We use the SCAPE dataset with 71 human shapes containing various large scale deformations [Anguelov et al. 2005]. As shown in Fig. 2, the sparsity localized deformation component method [Neumann et al. 2013] produces shrinkage artifacts because it uses Euclidean coordinates directly and cannot handle rotations well. The method [Huang et al. 2014] analyzing in the deformation gradient domain is better than [Neumann et al. 2013] in the leg region. However for the arm region with diverse large rotations, both of these existing methods fail to extract meaningful deformation components. Our method works well, extracting a basis with meaningful deformations. We further test the generalizability of our extracted basis. To achieve this, we split the SCAPE dataset into a training set and a test set. 36 shapes are randomly selected to form the training set and the rest are taken as the test set. As demonstrated in Fig. 3, the reconstruction error for the new shapes in the test set reduces with an increasing number of basis modes. Moreover, with our method the reconstruction error drops much more quickly than [Huang et al. 2014; Neumann et al. 2013].

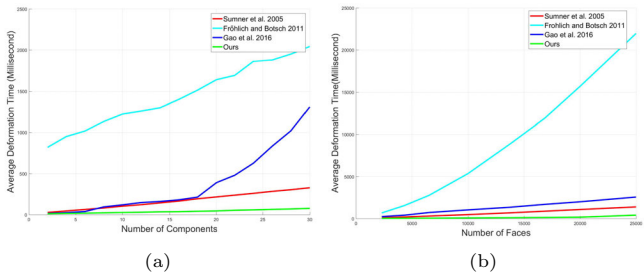

Fig. 6. Deformation running times of different methods w.r.t. increasing number of basis modes and triangle faces for the example in Fig. 18. (a) the running times w.r.t. an increasing number of basis modes with a fixed number (4K) of triangle faces, (b) the running times w.r.t. an increasing number of triangle faces with fixed (20) components. Our method scales well in both cases and is significantly faster than existing data-driven deformation methods, especially with larger numbers of basis modes and triangle faces.

|         | Faces | #. Basis Modes | Deformation Time (ms) |
|---------|-------|----------------|-----------------------|
| Fig. 14 | 2304  | 2              | 7                     |
| Fig. 16 | 2304  | 10             | 12                    |
| Fig. 18 | 4300  | 26             | 57                    |
| Fig. 1  | 6796  | 21             | 106                   |
| Fig. 20 | 4360  | 20             | 61                    |
| Fig. 22 | 4100  | 20             | 56                    |

Table 1. Statistics of the deformation running times.

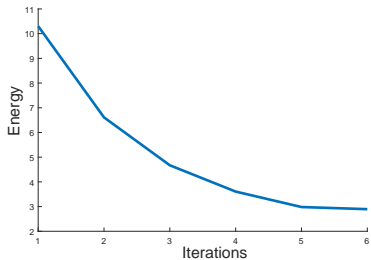

Fig. 7. Deformation energy over Gauss-Newton iterations for the example in Fig. 17. The energy monotonically decreases and converges over a few iterations.

To demonstrate how our sparse data-driven deformation method works, we show an example with synthetic basis modes, in the form of Gaussian function offset over a square shape, centered at a $5 \times 5$ grid. Figure 4 shows all the basis modes. In Fig. 5, we compare the deformation results with different methods. The method [Gao et al. 2016] (b) and our method without the sparsity term (c) tend to have a large number of basis modes involved to represent the deformed shape. Our method with the sparsity term (d) on the other hand prefers to use fewer basis modes if possible. The contribution of each basis mode and dominant modes are shown on the right. By using the sparsity promoting term, our method uses a smaller number of basis modes, leading to more localized data-driven deformation results.

**Time efficiency.** Our data-driven deformation is much more efficient and scalable than existing methods [Fröhlich and Botsch 2011; Gao et al. 2016; Sumner et al. 2005]. To evaluate time efficiency, we use SCAPE dataset simplified to $4K$ triangles with up to 30 deformation modes (or examples depending on the method) as the basis. Figure 6 shows the average deformation time w.r.t. to increasing numbers of deformation basis modes and mesh triangles, using different methods. Compared with the other three state-of-the-art data-driven methods, our method performs fastest, especially when the size of the basis is larger and/or the mesh contains more triangles. With the help of precomputation and parallelization, our method performs faster than [Sumner et al. 2005]. For [Fröhlich and Botsch 2011], since the Jacobian matrix is changing after each optimization iteration, it cannot be pre-decomposed, which slows down the computation. The computation time of [Gao et al. 2016] grows quickly with increasing size of deformation basis. This is because they compute the energy gradient with respect to weights numerically. Our approach is much more efficient as the gradients are calculated analytically. Table 1 shows the average running times for different examples in the paper. Our method achieves real-time deformation performance whereas alternative methods cannot cope with input such as SCAPE datasets with larger number of basis modes in real time. Figure 7 shows how the deformation energy changes over iterations, which is reduced monotonically and converges in several iterations.

We also report the *preprocessing* times for the results in Fig. 1 with 6796 triangle faces. These only need to be performed once. The as-consistent-as-possible optimization of Eqns. (4) and (6) takes 1.39s and 9.92s, respectively. The time for precomputation of Eqn. (15) takes 5.02s. The Cholesky decomposition of the matrix $\mathbf{A}$ takes 0.83s on average.

**Evaluation of our deformation representation.** Our novel representation is effective to represent shapes with large-scale deformations. Given two exemplar models (a cylinder and a cylinder rotated by five cycles) in Fig. 8, thanks to this representation, the interpolation and extrapolation results shown in Fig. 9 are correctly produced, even for such excessive deformations. Figure 10 shows an example of blending the Buddha models with multiple topological handles. Figure 11 shows an example of interpolating SCAPE models with substantial added noise. These examples demonstrate that our representation can cope with high-genus models and is robust to noise. In all of these examples, we linearly blend these two models with contributions of $1 - t$ and $t$ from these models.

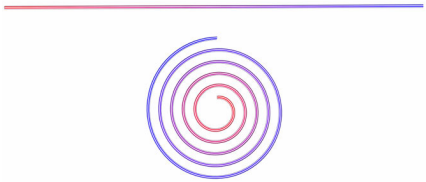

Fig. 8. The exemplar models with large-scale deformation: is a cylinder ($t = 0$), and the cylinder with five cycles of rotation ($t = 1$).

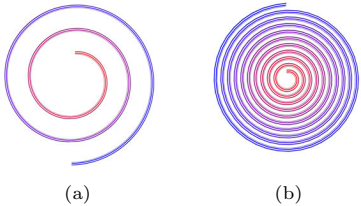

(a)       (b)

Fig. 9. The interpolation and extrapolation results of Fig. 8 which demonstrate that our representation can represent and blend very large-scale deformations. (a) the blended model with parameter $t = 0.5$, (b) the blended model with parameter $t = 2$.

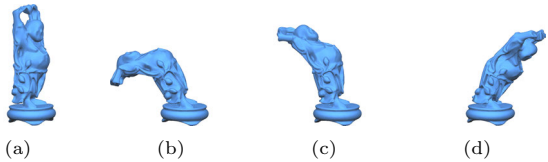

(a)     (b)     (c)     (d)

Fig. 10. Interpolation and extrapolation of high-genus models using our representation. (a)(b) two exemplar models to be blended, (c) the interpolated model with $t = 0.5$, (d) the extrapolated model with $t = -0.5$.

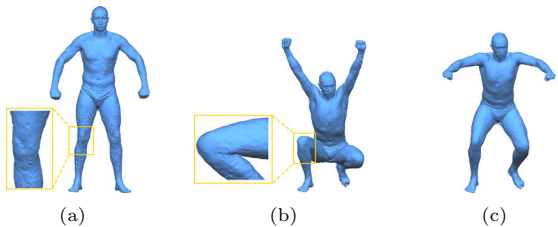

(a)      (b)      (c)

Fig. 11. Interpolation result for models with substantial Gaussian noise. (a)(b) noisy exemplar models, (c) the interpolated model.

**Evaluation of Initialization Strategies for Integer Programming.** We compare the breadth-first search (BFS) based initialization used in our implementation with trivial initialization (setting all $o_{k,i} = 1$ and all $\theta_{k,i} = 0$). With the help of the BFS-based initialization, the running time for axis orientation optimization in Eqn. 4 is greatly reduced. For the SCAPE dataset [Vlasic et al. 2008], the average running time of optimizing Eqn. 4 for each mesh is 14.80s with the BFS initialization while the running time is 544.89s with the trivial initialization. As explained in Sec. 3, BFS initialization has much less benefit for speeding up the optimization of Eqn. 6 where the average running time for each mesh model in the SCAPE dataset [Vlasic et al. 2008] is only reduced to 17.44s from 18.60s with the trivial initialization.

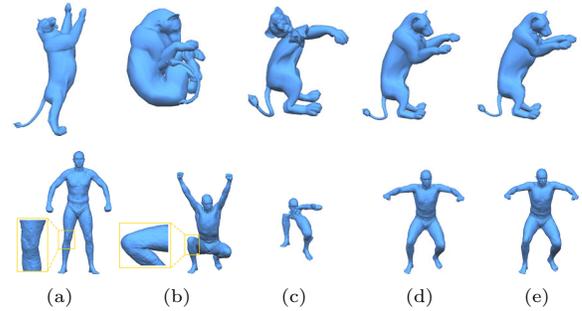

(a)     (b)     (c)     (d)     (e)

Fig. 12. Interpolation results for models with large scale deformations from [Lee et al. 2009] (the first row) and substantial added Gaussian noise (the second row). (a)(b) exemplar models, (c) the interpolated models with BFS initialization, (d) the interpolated models using the features obtained by the global integer programming with BFS initialization, (e) the interpolated models using the features from the global integer optimization by the trivial initialization.

Although the BFS initialization speeds up the convergence for optimization significantly, the initialization itself is not robust enough and the global optimization of Eqns. 4 and 6 are necessary. As shown in Fig. 12, for the shape with large scale deformations or substantial noise, the initial solution obtained using BFS initialization produces artifacts for the interpolated shapes (c), while after the global optimization the interpolated shapes are plausible (d). We also show that the final optimized results are not dependent on the initialization: the interpolated results are also correct with trivial initialization despite taking much longer time (e). Note that the times reported here are one-off for a given dataset.

**Comparison with state-of-the-art methods.** We now show several examples of our sparse data-driven deformation results and compare them with state-of-the-art methods. Figure 13 shows an example using the SCAPE dataset [Anguelov et al. 2005] as examples. All the green handles are fixed and the red handle is moved. The method [Gao et al. 2016] (b) does not produce satisfactory result because the basis modes are global and even local movement of one handle causes global deformation with obvious distortion on the arm. Our method without the sparsity term (c) produces more local result around the arm, due to the use of a localized basis. However, the deformation result is obtained with contributions of a large number of basis modes. Even if individual basis modes are spatially sparse, their overall effect can still involve unexpected global effect. In this case, it is clear that the knee is bent with no user indication of preference. Our method with the sparsity promoting term (d) produces a localized deformation result. The contributing basis modes are also visualized. Figure 14 shows different data-driven deformation results with two examples in its first row. Our rotation representation is able to handle very large rotations where the rod is twisted several times, which cannot be achieved using existing methods [Gao et al. 2016; Sumner et al. 2005; Wampler 2016]. The work [Wampler 2016] is not designed for large scale deformations. The result of [Wampler 2016] in the fifth row clearly shows that it performs poorly for exemplar models with large deformations. Moreover, the method [Wampler 2016] applies weights to the energy, and as such their weights must be non-negative. Our weights are applied in the gradient domain, where negative weights are not only acceptable, but also important, to allow extrapolation which is essential to fully exploit latent knowledge in the examples. Our $\ell_1$ sparse regularization minimizes an energy formulated in the $\ell_1$ norm for promoting to choose a minimum set of basis modes to produce plausible deformation, which has the benefits of avoiding overfitting, as demonstrated by various examples throughout the paper. This however is very different from [Wampler 2016], as their weights are non-negative and always summed to 1, so cannot promote sparsity.

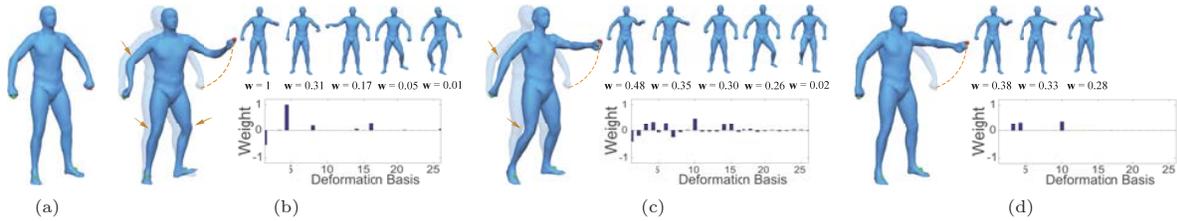

(a)  (b)  (c)  (d)

Fig. 13. Comparison of deformation results using the SCAPE dataset [Anguelov et al. 2005]. (a) input shape, (b) result of [Gao et al. 2016], (c) our result without the sparsity term, (d) our result with the sparsity term. We also visualize the contribution weight of each basis mode and the dominant basis modes. [Gao et al. 2016] uses global basis, whereas our method uses localized basis. By using sparsity regularization, our method produces deformation result with much fewer active modes, avoiding unintended global deformation produced by alternative methods.

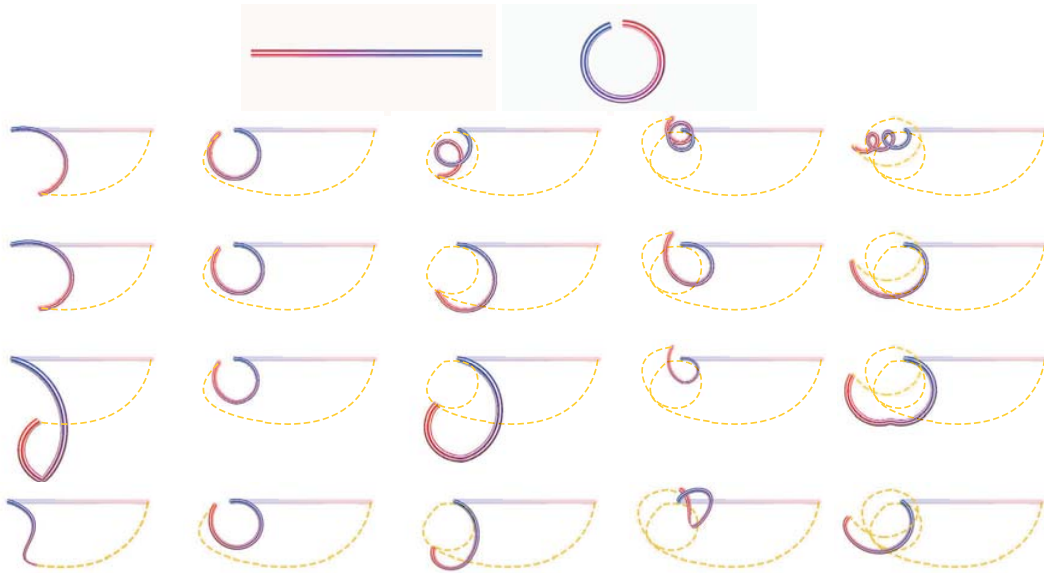

Fig. 14. Comparison of deformation results using the examples from the first row. Second row: our results, third row: [Gao et al. 2016], fourth row: [Sumner et al. 2005], fifth row: [Wampler 2016]. Please refer to the video. Thanks to our representation, our data-driven deformation method produces follows the handle movement and successfully generates deformed cylinder with multiple cycles of twisting, which cannot be produced by state-of-the-art methods.

Another example is shown in Fig. 16 with 10 example shapes given in Fig. 15. By avoid overfitting, our sparse data-driven deformation method produces smooth and intuitive deformation. The result of [Sumner et al. 2005] has significant artifacts because it cannot handle large rotations properly. For [Gao et al. 2016], the produced result has inconsistent cross sections caused by a large number of basis modes. This is less natural than the consistent ellipse shaped cross section produced by our method. We use faded rendering for the initial shape and a dashed line to visualize the handle movement. When a large number of exemplar models are used, such mixed exemplars are common, e.g. when humans with different body shapes and poses are included. This also occurs naturally when basis modes of multiple scales are considered simultaneously, as we will demonstrate later in the paper.

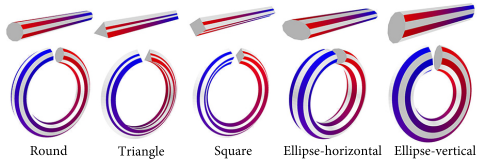

Fig. 15. Example cylinders with different styles.

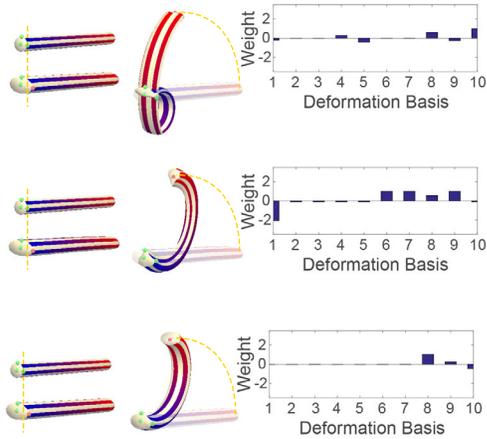

Fig. 16. Comparison of deformation results with examples from Fig. 15. First row: [Sumner et al. 2005], second row: [Gao et al. 2016], third row: our result. Please refer to the accompanying video. [Sumner et al. 2005] cannot handle large rotations thus produces distorted output. [Gao et al. 2016] uses a large number of basis which is underconstrained and generates inconsistent cross sections. Our method produces natural deformation result with consistent cross sections.

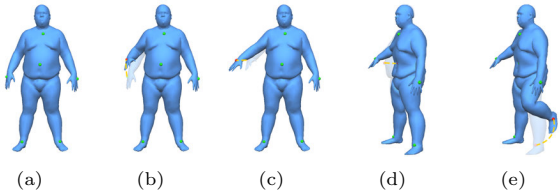

Fig. 17. A data-driven deformation sequence produced using our method. Natural local editing is generated without unintended deformation.

Figure 17 shows a sequence of deformed models with the MPI Dyna dataset [Pons-Moll et al. 2015] using our sparse data-driven deformation method. Even with a small number of handles, our sparse deformation result produces desired local editing as specified by user handles. As shown in Fig. 1, unexpected deformation occurs for the results of alternative methods, including turning of the head in the result of [Sumner et al. 2005] and the substantial movement of the right leg in the result of [Gao et al. 2016] despite no movement of related handles. Our method without the sparsity term produces less unexpected deformation due to the use of a localized basis, however, visible deformations also occur for the head and the right leg. Another example is shown in Fig. 18 using the SCAPE dataset [Anguelov et al. 2005]. Due to the larger rotations involved, existing data driven methods [Fröhlich and Botsch 2011; Sumner et al. 2005] and to a lesser extent [Gao et al. 2016] have artifacts of unnatural deformation. In the results of [Fröhlich and Botsch 2011; Gao et al. 2016], the left foot is turned with no movement of the related handle. Our method produces natural deformation and shapes are preserved for unmodified regions, thanks to the sparse deformation. Figure 19 compares our result with the results of [Wampler 2016]. Either with the global PCA basis or with the same spatially localized basis as our method, Wampler [2016] produce artifacts on the arm due to the limitation of handling large-scale deformations.

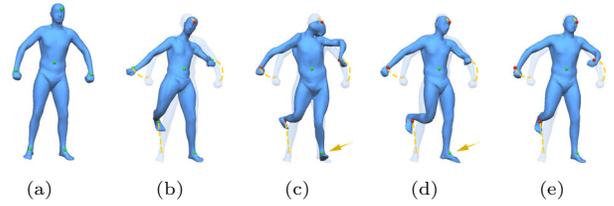

Fig. 18. Comparison of deformation results using the SCAPE dataset. (a) input shape, (b) [Sumner et al. 2005], (c) [Fröhlich and Botsch 2011], (d) [Gao et al. 2016], (e) our result. The methods [Fröhlich and Botsch 2011; Sumner et al. 2005] produce visible distortions due to large rotations involved in the deformation. Thanks to the sparsity regularization, our result suppresses unintended movement, which happen in the results of existing methods.

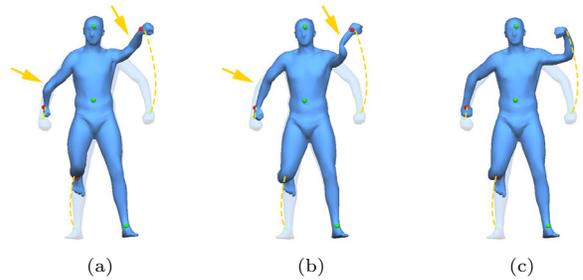

Fig. 19. Comparison of deformation results with [Wampler 2016] using the SCAPE dataset. (a) [Wampler 2016] with global PCA basis, (b) [Wampler 2016] with spatially localized basis, (c) our method.

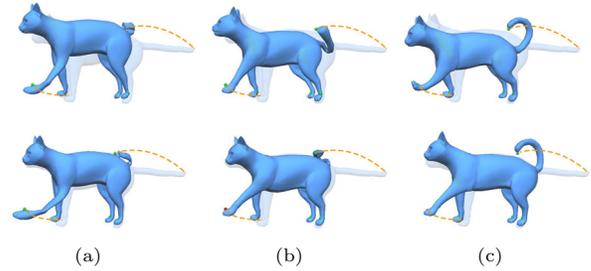

Fig. 20. Comparison of deformation results with methods with localized basis. (a) results of [Neumann et al. 2013], (b) results of [Huang et al. 2014], (c) our results. Top row: results without the sparse regularization term, bottom row: results with the sparse regularization term. The methods [Huang et al. 2014; Neumann et al. 2013] cannot handle large rotations well, whereas our method produces natural deformation. Using sparsity regularized (bottom row) effectively suppresses unintended deformation (top row).

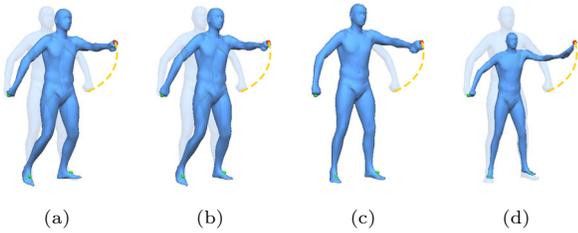

Fig. 21. Comparison of deformation results with different parameter $\lambda$: (a) $\lambda = 0$ (NZM = 18), (b) $\lambda = 0.05$ (NZM = 13), (c) $\lambda = 0.5$ (NZM = 3), (d) $\lambda = 20$ (NZM = 2). NZM refers to the number of basis deformation modes with non-zero weights. Stronger sparsity regularization is obtained with increasing $\lambda$. When $\lambda$ is too large, the deformation result favors sparsity over distortion and produces artifacts.

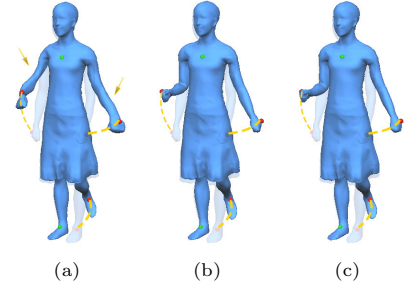

Fig. 22. Our sparse data driven deformation results with different number of basis modes $K$. (a) $K = 10$, (b) $K = 20$, (c) $K = 30$. With too few basis modes ($K = 10$), the method cannot cover plausible deformations and causes artifacts. With reasonably large $K$ (20 or 30), the results look plausible.

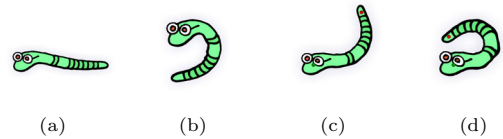

Fig. 23. 2D deformation using our method. (a)(b) exemplar models in the 2D space, (c)(d) deformed models with our method.

We compare our method with [Huang et al. 2014; Neumann et al. 2013] with localized bases in Fig. 20. The method [Neumann et al. 2013] uses Euclidean coordinates which cannot represent rotations correctly. As a result, the extracted basis deformations are inappropriate, leading to significant deformation artifacts. The method [Huang et al. 2014] based on deformation gradients can represent rotations but fail to handle large rotations. The method produces reasonable result for the paw movement, but obvious artifacts with the tail movement. Our method generates natural deformation. Although the original methods [Huang et al. 2014; Neumann et al. 2013] do not have the sparsity term we introduce in this paper, we demonstrate the effect of adding this regularization to each method. By incorporating this term, unexpected global movement is substantially suppressed.

**Parameters.** In our sparse data driven deformation, the parameter $\lambda$ is essential to control sparseness. Figure 21 shows the results of our method with different settings of $\lambda$. Note that in this example, the deformation constraint only involves moving an arm. With increasing $\lambda$, the sparse regularization becomes stronger, which leads to reduced global movement out of the region of interest, as well as the reduced number of basis deformation modes with non-zero weights (implemented by testing if the weight is $> 10^{-8}$). Figure 21 uses an extreme setting with $\lambda = 10$, which ends up with only 2 basis modes and severely distorts the shape. $\lambda = 0.5$ works well, and is used for all the other examples in this paper. It is clear that our algorithm is insensitive to the choice of $\lambda$. It has large changes when $\lambda$ changes by 10 times, but for values around 0.5 (0.4-0.6), it gives almost the same results. Another parameter is the number of basis modes $K$. Figure 22 shows an example of our sparse data-driven deformation with different $K$ values. It can be seen that when too few basis modes are used ($K = 10$) it produces artifacts as the basis is insufficient to cover plausible deformation space. When $K$ is sufficient large, high quality deformation results are produced. Increasing $K$ from 20 to 30 produces almost identical result, although with longer running time.

*More experiments.* Our algorithm can be directly applied to the deformation of 2D models. An example is shown in Fig. 23. We also show that our method works well for non-articulated models, such as deformation of faces (see Fig. 24) using the dataset from [Zhang et al. 2004] as well as garment (see Fig. 26 with exemplars shown in Fig. 25).

Figure 27 shows an example of our sparse deformation when the basis involves multiscale deformation modes, which are extracted by adjusting the range parameters (see [Neumann et al. 2013]). This demonstrates the capability of our sparse regularization: by using a most compact set of deformation modes to interpret handle movements, our method automatically selects suitable basis modes for both small-scale facial expression editing and large-scale pose editing (see also the accompanying video). We also use the multiresolution technique in this example to further improve deformation details.

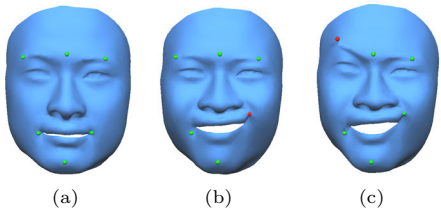

Fig. 24. Deformation of a 3D face model. (a) is the reference model to be deformed, (b) and (c) are the deformed models with our method.

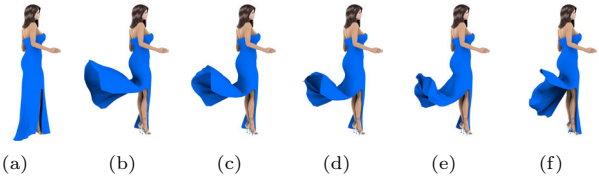

Fig. 25. Exemplar models for deformation of garment in Fig. 26.

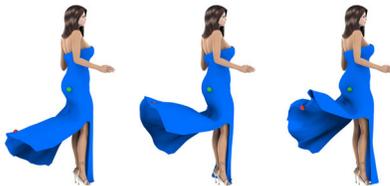

Fig. 26. Our deformation results of garment with exemplars in Fig. 25.

## 6 CONCLUSIONS

In this paper, we propose a simple and effective representation to handle large rotations, which is formulated as an as-consistent-as-possible optimization problem. This new representation has advantages of both efficient to compute and optimize, and can handle very large rotations (several cycles) where even recent existing rotation invariant methods fail. We further propose an approach to sparse data-driven deformation. By incorporating sparsity regularization, fewer essential basis modes are used to drive deformation, which helps to make the deformation more stable and produce more plausible deformation results. As realtime performance is essential for interactive deformation, we develop a highly efficient solution using pre-computation, which allows realtime deformation with larger size of basis than existing methods. Extensive experiments show that our method produce substantially better data-driven deformation results than state-of-the-art methods, suppressing unintended movement.

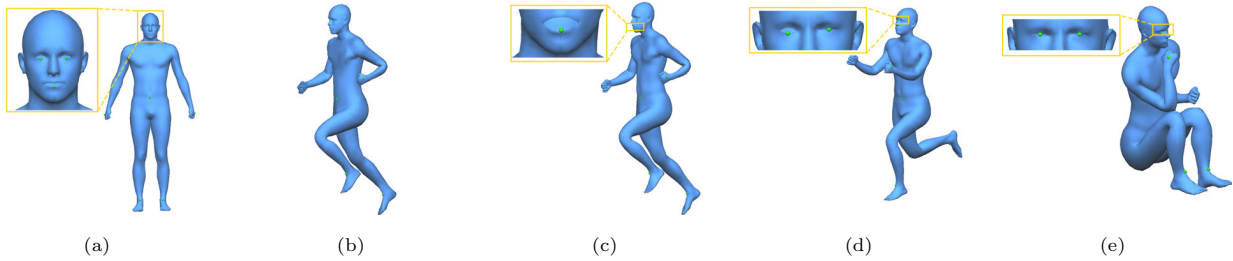

Fig. 27. Our sparse data driven deformation results using multiscale deformation modes as the basis. (a) is the reference model. (b) is the deformation result with the simplified mesh. (c)-(e) are the deformed results on the high resolution mesh with both facial and body deformation. The mouth of (c) is open, the left eye of (d) is closed, and both eyes of (e) are closed. Our method automatically selects suitable basis modes for both small-scale facial expression editing and large-scale pose editing.

## APPENDIX

**Efficient solution of Eqn. (15) using precomputation.**

Some computations in Sec. 4.3 can be pre-computed, making the algorithm much more efficient. Details are given below. For simplicity, let $\tilde{\mathbf{R}}_i(\mathbf{w}_t) = \exp(\sum_l w_{t,l} \log \tilde{\mathbf{R}}_{l,i})$. Then, terms in Eqn. (15) can be rewritten as:

$$\mathbf{T}_i(\mathbf{w}_t)\mathbf{e}_{1,ij} = \tilde{\mathbf{R}}_i(\mathbf{w}_t) \sum_l w_{t,l} \underline{\tilde{\mathbf{S}}_{l,i}\mathbf{e}_{1,ij}} \tag{28}$$

We use underscore for the term $\tilde{\mathbf{S}}_{l,i}\mathbf{e}_{1,ij}$ to indicate that it can be pre-computed before optimization to save time. The derivative term multiplied by the edge vector becomes:

$$\frac{\partial \mathbf{T}_i(\mathbf{w}_t)}{\partial w_l}\mathbf{e}_{1,ij} = \tilde{\mathbf{R}}_i(\mathbf{w}_t)(\sum_l w_{t,l} \underline{\log \tilde{\mathbf{R}}_{l,i} \tilde{\mathbf{S}}_{l,i}\mathbf{e}_{1,ij}} + \underline{\tilde{\mathbf{S}}_{l,i}\mathbf{e}_{1,ij}}) \tag{29}$$

One of the most time consuming step is to calculate $\sum_{i=1}^3 \mathbf{J}_i^T \mathbf{J}_i$. This calculation is equivalent to the following step:

$$(\sum_l w_{t,l} \underline{\mathbf{e}_{1,ij}^T \tilde{\mathbf{S}}_{l,i}^T \log \tilde{\mathbf{R}}_{l,i}})^T + \underline{\mathbf{e}_{1,ij}^T \tilde{\mathbf{S}}_{l,i}^T})\tilde{\mathbf{R}}_i(\mathbf{w}_t)^T \tag{30}$$
$$\times \tilde{\mathbf{R}}_i(\mathbf{w}_t)(\sum_l w_{t,l} \underline{\log \tilde{\mathbf{R}}_{l,i} \tilde{\mathbf{S}}_{l,i}\mathbf{e}_{1,ij}} + \underline{\tilde{\mathbf{S}}_{l,i}\mathbf{e}_{1,ij}})$$

With some derivation, this formulation can be further rewritten as:

$$(\sum_l w_{t,l} \mathbf{e}_{1,ij}^T \tilde{\mathbf{S}}_{l,i}^T \log \hat{\mathbf{R}}_{l,i}^T + \mathbf{e}_{1,ij}^T \tilde{\mathbf{S}}_{l,i}^T) \hat{\mathbf{R}}_i (\mathbf{w}_t)^T \quad (31)$$

$$\times \hat{\mathbf{R}}_i(\mathbf{w}_t) (\sum_l w_{t,l} \underline{\log \hat{\mathbf{R}}_{l,i} \tilde{\mathbf{S}}_{l,i} \mathbf{e}_{1,ij}} + \underline{\tilde{\mathbf{S}}_{l,i} \mathbf{e}_{1,ij}})$$

$$= \sum_l w_{t,l}^2 \mathbf{e}_{1,ij}^T \tilde{\mathbf{S}}_{l,i}^T \log \hat{\mathbf{R}}_{l,i}^T \underline{\log \hat{\mathbf{R}}_{l,i} \tilde{\mathbf{S}}_{l,i} \mathbf{e}_{1,ij}}$$

$$+ \sum_{l_1,l_2} w_{t_1} w_{t_2} \mathbf{e}_{1,ij}^T \tilde{\mathbf{S}}_{l_1,i}^T \log \hat{\mathbf{R}}_{l_1,i})^T \underline{\log \hat{\mathbf{R}}_{l_2,i} \tilde{\mathbf{S}}_{l_2,i} \mathbf{e}_{1,ij}}$$

$$+ \sum_l w_l \mathbf{e}_{1,ij}^T \tilde{\mathbf{S}}_{l,i}^T \log \hat{\mathbf{R}}_{l,i}^T \tilde{\mathbf{S}}_{l,i} \mathbf{e}_{1,ij}}$$

$$+ \sum_l w_l \mathbf{e}_{1,ij}^T \tilde{\mathbf{S}}_{l,i}^T \log \hat{\mathbf{R}}_{l,i} \tilde{\mathbf{S}}_{l,i} \mathbf{e}_{1,ij} + \underline{\mathbf{e}_{1,ij}^T \tilde{\mathbf{S}}_{l,i}^T \tilde{\mathbf{S}}_{l,i} \mathbf{e}_{1,ij}}.$$

The underlined terms do not change during the iterative optimization and can be calculated in advance, and only the remaining terms involving $\mathbf{w}_t$ need to be calculated. This makes the algorithm over 10 times faster (see the details in the experiments). Our precomputation significantly improves the efficiency. For the case with 30 deformation basis modes, each containing 4300 triangle faces, the size of the matrix $\mathbf{J}$ is $12900 \times 30$, where the number of rows is the number of half edges (i.e. three times the number of faces), and the number of columns is the number of deformation basis modes. Since $\mathbf{J}$ is a dense matrix, the time to calculate $\sum_{i=1}^{3} \mathbf{J}_i^T \mathbf{J}_i$ directly costs 38ms. For a typical scenario of five Gauss-Newton steps, it takes 190 ms, impeding real time performance. With precomputation, it only takes 1 ms for each Gauss-Newton step, sufficient for real-time deformation.

## ACKNOWLEDGMENTS

The authors would like to thank Kevin Wampler for his kind help for reproducing the results of [Wampler 2016] also thank Yu Wang for providing the mesh models in Fig. 23. This work was supported by the National Natural Science Foundation of China (No. 61502453 and No. 61611130215 ), Royal Society-Newton Mobility Grant (No. IE150731), CCF-Tencent Open Research Fund (No. AGR20160118), Knowledge Innovation Program of the Institute of Computing Technology of the Chinese Academy of Sciences (ICT20166040).